# Infrared Emission and Dynamics of Outflows in Late-Type Stars

Željko Ivezić and Moshe Elitzur

Department of Physics and Astronomy,
University of Kentucky, Lexington, KY 40506-0055

## ABSTRACT

The dynamical structure and infrared emission of winds around late-type stars are studied in a self-consistent model that couples the equations of motion and radiative transfer. Thanks to its scaling properties, both the dynamics and IR spectrum of the solution are fully characterized by $\tau_F$, the flux averaged optical depth of the wind. Five types of dust grains are considered: astronomical silicate, crystalline olivine, graphite, amorphous carbon and SiC, as well as mixtures. Analysis of infrared signatures provides constraints on the grain chemical composition and indications for the simultaneous existence of silicate and carbon grains. The abundances of crystalline olivine in Si-dominated grains and of SiC in C-dominated grains are found to be limited to $\lesssim$ 20–30%. Furthermore, in carbonaceous grains carbon is predominantly in amorphous form, rather than graphite. In mixtures, carbonaceous grains tend to dominate the dynamic behavior while silicate and SiC grains dominate the IR signature. The region of parameter space where radiation pressure can support a given mass-loss rate is identified, replacing the common misconception $\dot{M}v \leq L_*/c$, and it shows that radiatively driven winds can explain the highest mass loss rates observed to date. A new method to derive mass loss rates from IR data is presented, and its results agree with other determinations.

Theoretical spectra and colors are in good agreement with observations. IRAS LRS classes are associated with $\tau_F$ for various grain materials and the regions of color-color diagrams expected to be populated by late-type stars are identified. For a given grain composition, location in the color-color diagram follows a track with position along the track determined by $\tau_F$. We show that cirrus emission can severely affect point source measurements to the extent that their listed IRAS long wavelength fluxes are unreliable. Whenever the listed IRAS flag *cirr*3 exceeds the listed 60 $\mu$m flux by more than a factor of 2, the 60 and 100 $\mu$m fluxes are no longer indicative of the underlying point source. After accounting for cirrus contamination, essentially all IRAS point sources (95%) located in the relevant regions of the color-color diagrams can be explained as late-type stars. There is no need to invoke time dependent effects, such as detached shells, for example, to explain either the colors or mass loss rates of these sources. Although various indications of time varying mass-loss rates exist in numerous sources, the infrared properties of this class of stars are well explained as a whole with steady state outflows.

*Subject headings:* stars: late-type stars — stars: circumstellar shells — stars: mass-loss — stars: infrared radiation — circumstellar matter: dust

## 1. INTRODUCTION

The radiation of many astronomical objects undergoes significant processing by surrounding dust. Since this processing usually results in a shift to the infra-red, most of the observational information is available in that regime. Recent progress in observational capabilities makes the IR signature of such objects a powerful tool of analysis. However, detailed radiative transfer calculations are essential for a meaningful interpretation of the observations and such calculations frequently require ad hoc assumptions about the spatial distribution of dust around the central source.

Late-type stars are a notable exception. These stars are losing mass at significant rates, up to $\sim 10^{-4}$ $M_\odot$ yr$^{-1}$. It is widely believed that, once dust grains form at the tops of the extended atmospheres, radiation pressure on the grains is the driving force of further outflow. Therefore, rather than assumed, the dust distribution can be actually calculated from a self-consistent solution of the wind dynamics and radiative transfer. A full analysis of this coupled problem was presented recently by Netzer & Elitzur (1993, hereafter NE) and the current study is a continuation of that work. We present a thorough investigation of the dynamics and spectra of dusty envelopes around giant and supergiant stars and their implications for observations, especially IRAS data. Since the results are derived from a self-consistent model they provide a reliable indication of both the capabilities and limitations of IR data.

The model and input parameters are briefly described in §2. Dynamical results are presented in §3, spectral properties in §4 and in §5 we show how to determine outflow parameters (including $\dot{M}$) from IRAS observations. The results are summarized and discussed in §6.

## 2. THE MODEL AND INPUT PARAMETERS

A detailed description of the model and its equations is provided in NE. Briefly, the circumstellar envelope, which consists of gas and dust, is treated as a two-component fluid to account for relative drifts. The equations of motion for the two components are solved as a set coupled to the radiative transfer equation, treated in the moment form. The only difference from NE (see their eq. 22) is that the function $q$, introduced by Adams & Shu (1986) to facilitate convenient closure of the moment equations, is parametrized as

$$q_\nu = 2 \exp\left[-\alpha(\tau_\nu^T - \tau_\nu)\right]. \tag{2-1}$$

Here $\tau_\nu^T$ is the overall optical depth at frequency $\nu$ and $\alpha$ ($\sim 1$) is a dimensionless parameter, determined from flux conservation. This form provides a more adequate numerical scheme than the one used in NE. Model input are the stellar parameters (luminosity $L_*$, effective temperature $T_*$ and mass $M_*$), the mass-loss rate $\dot{M}$ and the optical properties of the dust grains, specified by absorption and scattering opacities per unit mass. The model produces the radial profiles of the density, outflow velocity and dust temperature, as well as the radiation field in the envelope.

Similar to NE, the stellar luminosity is varied from $10^3$ $L_\odot$ to $10^5$ $L_\odot$, the full range of giant and supergiant luminosities. A single effective temperature, $T_* = 2500$ K, is used for all the results presented below. Varying $T_*$ in the range 2000–3000 K has a minor impact on the solution through changes in the spectral distribution of the stellar radiation. This small effect is noticeable only in thin envelopes, since the stellar radiation is entirely reprocessed in thick envelopes, and observed IR spectral colors are practically independent of $T_*$. The stellar mass is varied from 0.5 $M_\odot$ to 10 $M_\odot$ while $\dot{M}$ is varied in the range $10^{-7}$— $5 \times 10^{-4}$ $M_\odot$ yr$^{-1}$, covering the entire observed range (e.g. Knapp et al, 1982; Wood et al, 1992; NE).

The optical properties of the dust grains are covered more completely than in NE. Infrared spectra indicate that dust grains can have various chemical compositions and in this work we consider the five major grain types that have been proposed for late-type stars. Dielectric coefficients for "astronomical silicate" and graphite are taken from Draine & Lee (1984) and Draine (1987). The resulting absorption efficiency $Q_{abs}$ for silicate grains was slightly modified in accordance with the IRAS data analysis of Simpson (1991), which indicates that the strengths of the silicate peaks, especially the one at 18 $\mu$m, should be enhanced. Efficiencies for crystalline olivine are from Aannestad (1975) and Koike et al. (1993). Dielectric coefficients for amorphous carbon are from Hanner (1988), for silicate carbide from Pégourié (1988). Although the presence of polycyclic aromatic hydrocarbons (PAH) in the envelopes of late-type stars has been suggested (Buss et al, 1991; Cherchneff et al, 1992), we do not include those; even if present, PAH's do not affect significantly the outflow dynamics (Cherchneff et al, 1991). Absorption and scattering efficiencies are evaluated using the Mie theory, assuming the dust grains are homogeneous spheres.

Figure 1 displays $Q_{abs}$ for randomly oriented spheres (graphite has an anisotropic dielectric constant). Astronomical silicate has characteristic peaks at 9.7 $\mu$m and 18 $\mu$m, SiC at 11.3 $\mu$m, crystalline olivine at both 9.7 and 11.3 $\mu$m, as well as a rich structure between 20 and 40 $\mu$m. The effect of the Simpson modifications to the Draine & Lee calculations can be gauged from the two curves drawn for silicate; the one employed in subsequent analysis is the thick solid line. Graphite shows a broad bump at $\sim 20$ $\mu$m, while $Q_{abs}$ for amorphous carbon is a featureless, monotonic function of wavelength. In the long-wavelength region ($\lambda \gtrsim 60$ $\mu$m), $Q_{abs}$ for graphite and other crystalline grains is proportional to $\lambda^{-2}$, while for SiC and amorphous carbon and silicate $Q_{abs} \propto \lambda^{-1.5}$, a signature of more disordered material. While Draine & Lee employed the former for silicate grains, both Rowan-Robinson et al. (1986) and Hashimoto et al. (1990) advocate the latter. Our results provide a slightly better agreement with observations for amorphous silicate, so this is the form we use.

In addition to pure grain material, we consider mixed compositions using the simple rule

$$\kappa_{\mathrm{mix}} = \frac{p_1 \kappa_1 + p_2 \kappa_2}{p_1 + p_2} \qquad (2\text{-}2)$$

for a mixture in the variable proportion $p_1$:$p_2$ of two components with opacities $\kappa_1$ and $\kappa_2$, respectively. Mixture computations are performed with a three-fluid hydrodynamics scheme because the different components of a mix can drift at different velocities. We find that in general, such calculations do not produce significant differences in either outflow velocities or emerging spectra from those of two-fluid calculations in which the dust is treated as a single component with the opacity $\kappa_{\mathrm{mix}}$.

The grain material affects not only the opacity but also the dust condensation temperature, which determines $r_0$, the inner radius of the circumstellar dust shell. Keady, Hall and Ridgway (1987) point out that in mixtures, different components will condense at somewhat different radii because of the differences in condensation temperature. Following Bedijn (1987) we have varied the dust condensation temperature from 700 K to 1000 K and found that it has a minor impact on the solution. For simplicity, a single temperature of 850 K is used in all subsequent analyses.

Grain sizes have been analyzed by NE, who conclude that they are mostly of order 0.05 $\mu$m, in agreement with previous models (Rowan-Robinson and Harris, 1982, 1983ab; Griffin, 1990; Harvey et al, 1991). We adopt this as a single size for all grains. Because of the scaling properties of grain opacities, grain size variations within reason are irrelevant at wavelengths above $\sim 5\mu$m.

Finally, $r_{gd}$, the initial gas to dust ratio prior to drift effects, has a significant impact on the dynamics. The outflow terminal velocity is roughly proportional to the $1/r_{gd}$, a relation we have verified in the range $100 < r_{gd} < 500$. Volk et al (1992) assume $r_{gd} \sim 260$–$330$ for graphite and SiC grains, Knapp (1985) prefers 200 for oxygen stars and 670 for carbon stars while Herman et al (1986) find $250 \pm 100$ for oxygen rich Miras and $160 \pm 40$ for OH/IR stars. All the results presented here are obtained with $r_{gd} = 200$. This value is essentially the same as the one proposed by Jura (1986) and it produces terminal velocities in agreement with observations.

## 3. DYNAMICAL PROPERTIES

For a given mass-loss rate, stellar luminosity and dust opacity, the outflow velocity $v$ obeys

$$\dot{M}v = \tau_F \frac{L_*}{c}, \tag{3-1}$$

where

$$\tau_F = \frac{1}{F} \int_0^\infty \tau_\nu F_\nu d\nu \tag{3-2}$$

is the flux averaged optical depth and $F = \int F_\nu d\nu$ is the overall flux. This result, valid when the gravitational pull of the star is negligible and $\frac{1}{2}\dot{M}v^2 \ll L_*$ (see Appendix A), was derived long ago by Salpeter (1974b, with a misprint of factor 2) and has generated a considerable amount of confusion in the literature ever since. It has often been stated that $\tau_F$ must be less than or at most equal to 1 because of momentum conservation, so that a wind with $\dot{M}v \gg L_*/c$ cannot be radiatively driven (e.g. Zuckerman & Dyck, 1986). However, although momentum conservation appears to be violated for $\tau_F > 1$, this is only an apparent conflict, resolved by the difference between the velocities of matter and radiation (Appendix A). Indeed, certain models we have computed produce values of $\tau_F$ as large as 20, so $\dot{M}v$ can exceed $L_*/c$ by this much in radiatively driven winds in late-type stars. Therefore, the detection of winds with $\dot{M}v \simeq 10 L_*/c$ does not rule out radiation pressure as their driving force, as argued by Zuckerman & Dyck (1986), nor is there any need to invoke time dependent episodes of mass-loss rate and/or luminosity, as suggested by Jura (1986).

What is the actual phase-space domain of radiatively driven winds? Taking into account the stellar gravitational attraction, equation 3-1 becomes

$$\dot{M}v = \tau_F \frac{L_*}{c}(1 - \Gamma^{-1}), \tag{3-3}$$

where

$$\Gamma = \frac{L_* \langle \chi_F \rangle}{4\pi G c M_*} \tag{3-4}$$

is the ratio of radiative to gravitational forces in the envelope. Here

$$\langle \chi_F \rangle = \frac{\tau_F}{\int_{r_0}^\infty \rho(r) dr} = \frac{\int_{r_0}^\infty \chi_F(r) \rho(r) dr}{\int_{r_0}^\infty \rho(r) dr} \tag{3-5}$$

is the spatial average, weighted by the normalized density profile $\rho/\int \rho dr$, of $\chi_F(r)$, the flux-averaged opacity per unit mass (cf eq. 3-2). Therefore, rather than $\tau_F < 1$, the domain of radiatively driven winds is determined by $\Gamma > 1$, or

$$\frac{M_*}{L_*} < \frac{\langle \chi_F \rangle}{4\pi G c}. \tag{3-6}$$

This relation is similar to the Eddington luminosity condition, only the relevant opacity is averaged over frequency and space. Numerically, this condition is $M/L_4 < 0.77 \langle \chi_F \rangle$, where $M = M_*/M_\odot$, $L_4 = L_*/10^4 L_\odot$ and $\langle \chi_F \rangle$ is in cm$^2$ g$^{-1}$. An order-of-magnitude estimate for $\langle \chi_F \rangle$ can be obtained from a calculation of the flux-averaged opacity at $r_0$, the dust formation point. If $\chi_F(r)$ were independent of position in the wind, $\langle \chi_F \rangle$ would be equal to $\chi_F(r_0)$. In reality, two effects cause $\chi_F$ to vary in the circumstellar shell introducing a dependence of $\langle \chi_F \rangle$ on $\dot{M}$, and in general $\langle \chi_F \rangle < \chi_F(r_0)$. Dust drift through the gas reduces the dust-to-gas ratio, hence the opacity per unit mass. When the mass loss rate decreases, $\chi_F$ decreases too because of the increased drift velocity. This effect disappears at higher mass loss rates when the drift decreases below the outflow velocity and the dust-to-gas ratio becomes constant. The other effect is the reddening of the radiation, which increases with the wind's column density, i.e., $\dot{M}$. Reddening shifts the weighting in the flux average toward longer wavelengths, where the opacity is generally declining. For carbonaceous grains the effect of any spectral features is always negligible and $\langle \chi_F \rangle$ decreases when $\dot{M}$ increases in optically thick ($\tau_F > 1$) winds. As a result, starting from very low mass loss rates $\langle \chi_F \rangle$ increases with $\dot{M}$ to a maximum, obtained when $\tau_F \lesssim 1$ and both drift and reddening are negligible, and declines with further increase in $\dot{M}$. Silicate grains, thanks to their prominent absorption peaks, are capable of capturing reddened radiation for all realistic mass loss rates and $\langle \chi_F \rangle$ is rising with $\dot{M}$ even in optically thick winds, at an enhanced rate reflecting the effect of the 9.7 $\mu$m feature. For all grains, models corresponding to typically observed parameters produce $\langle \chi_F \rangle \lesssim 30$ cm$^2$ g$^{-1}$.

Even when $\Gamma > 1$, radiation pressure can sustain the mass loss rate fed-in at the dust formation point only when the value of $v$ obtained from eq. 3-1 exceeds the thermal speed. We find this always to be the case. From the detailed model calculations we also find that $\langle \chi_F \rangle$ is a nearly unique function of $\dot{M}/L_*$, enabling us to identify the domain of radiatively driven winds as a region in the ($\dot{M}/L_*$, $M_*/L_*$) plane, displayed in figure 2. Results are presented for the various pure grain materials we consider and some representative mixtures. SiC is included only in a mixture with amorphous carbon since we find that observations rule out pure SiC grains as a single source of opacity (see §4 below). For a given grain material, radiation pressure can sustain winds only when their parameters fall in the region denoted with $\Gamma > 1$, bounded by the curve corresponding to that particular grain composition. The shape of each boundary reflects directly the dependence of $\langle \chi_F \rangle$ on $\dot{M}$. Segments at lower mass loss rates correspond to optically thin winds and are controlled by dust drift, those extending to high mass loss rates correspond to optically thick winds and are determined by the reddening of the radiation. It should be noted that the boundaries scale with the dust-to gas ratio; higher dust abundances shift the boundaries to the right, so that a mass-loss rate can be supported at larger values of $M_*/L_*$.

Figure 2 identifies the combinations of parameters that can sustain radiatively driven winds and replaces the erroneous relation $\dot{M}v \leq L_*/c$. Sources marked in the figure with stars have $\dot{M}v > L_*/c$ and were considered evidence against radiation pressure as the driving force of the winds (Zuckerman & Dyck, 1986; Jura, 1986). Instead, the figure shows that they all fall in the $\Gamma > 1$ region for carbonaceous grains; indeed, all are carbon stars. The locations of these stars were derived assuming that they all have $M_* = 10 M_\odot$, placing them as close as possible to the boundary of the forbidden region $\Gamma < 1$. Smaller values of $M_*$, which are more likely, move these stars away from the boundary deeper into the allowed region.

Van der Veen & Rugers (1989) find that, although the ratio $\dot{M}vc/L_*$ can reach $\sim$20–30 in oxygen rich stars, it never exceeds $\sim$15 in carbon stars. Employing equation 3-3, this observation is readily explained by the domains presented in figure 2 for the different grains.

The dynamical character of winds driven by radiation pressure on carbonaceous grains is quite different from that of winds that contain pure silicate grains. In mixtures, silicate grains have little impact on the dynamics. Even mixtures that contain as much as 50% silicates produce winds essentially identical to those of pure carbonaceous material. *Whenever present, carbonaceous grains tend to dominate the dynamics of the wind.* For a given luminosity and mass-loss rate, carbonaceous grains support radiatively driven winds at higher stellar masses than silicate grains as long as $\dot{M}/L_4 < 10^{-5}$, at which point the situation reverses.

## 4. SPECTRAL PROPERTIES

### 4.1. Scaling

For a given grain model, the effects of radiative transfer on the shell dynamics enter through the flux averaged $\tau_F$ and $\langle \chi_F \rangle$. Only the spectral shape

$$f_\nu = F_\nu/F \tag{4-1}$$

enters into the averaging procedure; the overall luminosity is irrelevant. An important scaling property of the radiative transfer problem is that, for a given grain composition, $f_\nu$ itself is a unique function of $\tau_F^T$, the overall flux-averaged optical depth. As a result, models with different sets of $L_*$, $M_*$ and $\dot{M}$ will have identical emerging spectra scaled with $L_*$ if they produce the same $\tau_F^T$. This remarkable scaling, briefly noted by Chan & Kwok (1990), enables us to present spectral shapes of the emerging radiation in terms of the single parameter $\tau_F^T$.

The reason for this scaling is quite simple. Because of the spherical symmetry, the radiative transfer equation is

$$\cos\theta \frac{\partial I_\nu}{\partial r} - \frac{\sin\theta}{r}\frac{\partial I_\nu}{\partial \theta} = \kappa_\nu(r)(S_\nu - I_\nu), \tag{4-2}$$

subject to a boundary condition that specifies the intensity at an inner radius $r_0$, the dust formation point. This radius is determined by the temperature of dust formation. Since the source function $S_\nu$ $(= B_\nu(T_d))$ is determined from an integral of the intensity itself through flux conservation, a given function $\kappa_\nu$ defined for all $r \geq r_0$ uniquely specifies a solution. With the dimensionless radial coordinate $x = r/r_0$, the radiative transfer equation becomes

$$\cos\theta \frac{\partial I_\nu}{\partial x} - \frac{\sin\theta}{x}\frac{\partial I_\nu}{\partial \theta} = \tau_\nu^T g(x)(S_\nu - I_\nu) \tag{4-3}$$

with its boundary condition specified at $x = 1$. Here $\tau_\nu^T$ is the overall optical depth of the system at frequency $\nu$, whose frequency variation is fully prescribed by the grain model, and $g(x)$ is the normalized dust density profile $\rho_d/\int \rho_d dx$ expressed in terms of the dimensionless variable $x$. A given profile $g(x)$ uniquely determines a family of solutions for $I_\nu$ whose members are characterized by their overall optical depth $\tau_\nu^T$. Models with the same $\tau_\nu^T$ have fluxes which are identical, except for an overall scaling with $L_*$ because of flux conservation. The same spectral shape preserves the values of $\tau_F$ (eq. 3-2) and $\langle \chi_F \rangle$ (eq. 3-5), which in turn determine the velocity profile (eq. 3-3), and through the mass conservation relation the density profile itself. Therefore, two models with different parameters but with the same $\tau_F^T$ will have the same profile $g(x)$, thus producing identical solutions only their fluxes will be scaled by the luminosities.

Scaling implies that for a given dust chemical composition, the distance-invariant spectral shape $f_\nu$ is a family of functions that depend on a single parameter, $\tau_F^T$, rather than the numerous parameters that characterize individual late-type stars. Therefore, unique correlations are expected between any pair of normalized spectral quantities for all stars; the distributions of such quantities (for example colors, strengths of spectral features) should display tight correlations rather than random scatter. At any given frequency, $f_\nu$ can be directly obtained from the data by considering observed normalized fluxes $F_\nu/F$, where $F$ is the bolometric flux. Figure 3a presents the distribution of normalized fluxes at 12 and 60 $\mu$m for a sample of 89 stars, listed in table 1. These are all the IRAS objects identified as late-type stars for which we were able to find listings in the literature for both total fluxes, luminosities, terminal velocities and mass-loss rates. Stars that are either cirrus contaminated (see §4.3 below) or have low quality IRAS fluxes are not included. This sample is used below for comparison with model calculations of mass-loss rates (§5.2). As can be seen from the figure, the normalized fluxes do display a tight correlation as expected, a correlation that holds over the entire observed range. Figure 3b displays the corresponding distribution of $(\nu F_\nu)_{60}/F$ and [25]–[12] color. The expected correlation is again evident, and can be used to estimate the total flux $F$ from IRAS data.

For comparison of our calculations with the data, computed model spectra are convoluted with the IRAS instrumental band profiles (Neugebauer et al, 1984; Bedijn, 1987) to produce the appropriate fluxes at $\lambda = 12, 25, 60$ and $100$ $\mu$m; colors for any pair $(\lambda_a, \lambda_b)$ are then calculated according to

$$[\lambda_a] - [\lambda_b] = \log \frac{F_\nu(\lambda_a)}{F_\nu(\lambda_b)}. \tag{4-4}$$

As is evident from figure 3, there is a close agreement between the data and the curves obtained from our model calculations for various relevant chemical compositions (SiC is only displayed in a 1:4 mixture with amorphous carbon because pure SiC has a too strong 11.3 $\mu$m feature; see below). The agreement between model predictions and observations is better than factor 2 in almost all cases. In addition, O-stars and C-stars clearly separate and congregate in accordance with the trend of the model curves for corresponding chemical compositions. This behavior is particularly prominent in panel b.

### 4.2. The Emerging Spectra

The six panels of figure 4 display a series of spectra for silicate dust with increasing $\tau_F$[1], as indicated, over the wavelength band of the IRAS Low Resolution Spectrometer (LRS). Dashed lines represent the direct stellar radiation emerging from the system, dotted lines are the dust emission and full lines are the total emission — the model predictions of observed spectra. In optically thin envelopes ($\tau_F < 1$) the strength of the 9.7 $\mu$m emission feature increases with optical depth until $\tau_F$ reaches $\sim 1$. With further increase in $\tau_F$, this emission turns into an absorption feature whose depth is proportional to $\tau_F$, as noted by Kwan and Scoville (1976). As an illustration, the inset in each panel displays the observed spectrum of an actual source whose IRAS number is indicated above the top left corner. Each panel also indicates the [25]–[12] color of the computed model while the number listed in large type in the inset is the IRAS color of the displayed source. The latter is obtained from measurements with the appropriate bandpass detectors, independent of the LRS instrument that produced the spectrum. The close agreement of spectral evolution between model predictions and observations is evident.

Because of the inherent strength of the silicate 9.7 $\mu$m feature, radiative transfer effects are important even in optically thin envelopes. As a result, deducing the dust contribution from the observations without a full radiative transfer calculation can lead to severe errors. Little-Marenin and Little (1990) and Stencel et al. (1990) attempted to determine the dust absorption coefficient by subtracting a 2500 K black-body emission from the observed spectrum, assuming that at 8 $\mu$m the dust contribution can be neglected. Some of their results are displayed in the left panels of figure 5, which show spectral features termed Sil, Sil+ and Sil++ by these authors. They propose that such features can be explained by annealing and aging of the dust grains, a proposal supported by the laboratory experiments of Nuth and Hecht (1990). Unfortunately, even for the thinnest envelopes the dust contribution is not negligible at 8 $\mu$m; already at $\tau_F = 0.1$, the dust emission at 8 $\mu$m is as strong as the stellar radiation (the corresponding optical depth at 9.7 $\mu$m is 0.3). The right panels of figure 5 display spectra obtained with *the same absorption coefficient*, that of astronomical silicate; only $\tau_F$ varies between the different models. Although the possibility of dust aging cannot be ruled out, optical depth effects provide a much simpler explanation for the observed overall spectral evolution. Our results are in agreement with the finding that the [25]–[12] color increases from the proposed Sil++ class to the Sil class; as can be seen from the figure, this sequence corresponds to an increasing $\tau_F$ and, hence, [25]–[12] color. The fundamental change in the spectral shape is caused by radiative transfer effects; major part of the alleged structure in the absorption coefficient simply reflects the spectral evolution of the emergent radiation when the dust contribution becomes comparable to the stellar contribution across the 9.7 $\mu$m feature. The minor secondary bump at $\sim$ 11.3 $\mu$m could reflect a small percentage of either crystalline olivine, aluminum or SiC grains. While a detailed fit for the spectra was not attempted, the two thin lines in the lower right panel display the dust and total emissions including a 20% crystalline olivine contribution, providing an adequate match for the observed spectrum of RU Her. As these results indicate, a full radiative transfer treatment cannot be avoided even for envelopes with small optical depths. Conclusions derived from the much simplified procedure of separating directly the stellar and shell contributions unfortunately cannot be considered reliable.

---

[1] In all the figures and subsequent discussion we use $\tau_F$ for $\tau_F^T$.

Graphite and amorphous carbon grains produce featureless spectra, similar to each other. The only difference between them is that graphite has a weak, broad feature between 20 $\mu$m and 40 $\mu$m, resulting in a somewhat flatter spectrum in this wavelength range. In mixtures, whenever silicate grains are present they tend to dominate the spectrum with their features. The spectral distribution resembles that of pure silicates and exhibits a similar dependence on $\tau_F$ even when the fraction of silicate grains is as low as 10%. While carbonaceous grains dominate their dynamics, *the infrared signature of mixtures is heavily dominated by silicate grains.* Infrared spectra of stars with dusty envelopes cannot discriminate well between a large range of silicate and carbonaceous grain mixtures.

### 4.3. Color-Color Diagrams

Location in IRAS color-color diagrams has become a widely used indicator of the nature of a source. In particular, beginning with Olnon et al. (1984), late-type stars were believed to occupy a region around the line corresponding to black-body emission with temperature $\gtrsim$ 100 K in the [60]–[25] — [25]–[12] color-color plane. Aided by LRS spectra, van der Veen & Habing (1988) refined the analysis and identified the appropriate IRAS region for late-type stars as [60]–[25] < 0 and [25]–[12] < 0.6, which we will refer to as the *VH-window*. In figure 6 we present the color-color diagrams for all IRAS sources whose colors fall in this window and with the highest quality (3) of measured flux in the first three bands and qualities 2 and 3 at the 100 $\mu$m band. This sample includes 1437 sources. The upper panel shows the [60]–[25] vs. [25]–[12] diagram, with our model predictions for selected grain compositions displayed as tracks. Position along each track is determined by $\tau_F$ and all tracks originate from the same spot, $\tau_F = 0$, corresponding to a black-body spectrum at 2500 K convoluted with the IRAS instrumental profiles. Indeed, the concentration of IRAS objects around this point is believed to represent stellar photospheres (Habing 1987). Distance from this common origin along each track increases with $\tau_F$ and the vertical marks delineate $\tau_F = 0.1$, 1.0 and 10.0. As $\tau_F$ increases, the dust emission becomes more prominent and the tracks of different grains branch out. The track for pure silicates resembles the color sequence from classical Mira variables to OH/IR stars proposed by Olnon et al. (1984).

The distribution displays a gap between the concentration of objects corresponding to stellar photospheres and those with winds dominated by silicate grains. Habing (1987) proposed an explanation for this gap, based on discontinuities in the mass-loss rate distribution. However, our models show that the gap can arise because silicate grains have a very steep dependence of [25]–[12] color on $\tau_F$ at small values of $\tau_F$. As can be seen from the figure, a source with $\tau_F = 0.1$ is already located beyond the gap, thus the gap is produced even for a uniform distribution of mass-loss rates. For carbonaceous grains other than graphite, $\tau_F = 0.1$ occurs much closer to $\tau_F = 0$, indeed the gap disappears in the color-color region where their tracks dominate. A huge gap would exist in that region if the dust were purely graphite, indicating that graphite cannot be a major component. This conclusion is strengthened by spectral analysis presented below.

The color-color diagram of IRAS objects limits the fraction of SiC in mixtures with amorphous carbon grains to less than 50%. The SiC feature at 11.3 $\mu$m is so prominent that it shifts the tracks for high SiC abundances to the far left of the color-color diagram ([25] − [12] < −0.9) where there are no IRAS sources. The analysis of LRS spectra described below also shows that the most plausible range for SiC abundance is ∼ 20–30%. These values produce colors in agreement with the IRAS observations and form the characteristic C-shape in the color-color diagram noted by Chan & Kwok (1988). Contrary to the proposals by Le Bertre (1988) and Chan & Kwok (1990), pure SiC grains are definitely ruled out as a plausible model for the dust. Similarly, the abundance of crystalline olivine in a mixture with silicate cannot exceed ∼ 20–30%. The lowest track displayed in the top panel, corresponding to 20% crystalline olivine, is already at the edge of the data distribution. Increasing further the abundance of this component produces tracks in a region devoid of data points.

The tracks for purely Si- and C-based grains outline the distribution boundaries of most IRAS sources in the VH-window. A significant separation exists between the track for astronomical silicate and those for all carbonaceous grains. However, the data points fill the entire region between these tracks and if this spread is real, it requires a similar spread in some property of the dust grains; populations of pure grains with distinct oxygen or carbon based chemistry cannot fill the whole region. Some continuous variation of grain properties is required to provide the spread of tracks in the color-color diagram indicated by the data.

One possible variation involves the long wavelength behavior of $Q_{abs}$ for silicate grains. The index $\beta$ of the power law $Q_{abs} \propto \lambda^{-\beta}$ could be smaller than the value 1.5 which we use and perhaps vary in the range $1 \leq \beta \leq 1.5$. Such a variation would affect only long wavelength fluxes and would raise the corresponding parts of color-color tracks, filling the required region. Also, even with a fixed value $\beta = 1.5$, the starting wavelength for this behavior could be shorter than the 60 $\mu$m which we assume and possibly vary continuously all the way to the 18 $\mu$m peak, resulting in a similar effect. Either modification produces tracks that fill most of the region bracketed by pure chemical compositions and mimic the distribution of IRAS sources. However, both possibilities suffer from similar difficulties with regard to two other aspects of the data. First, such modifications cannot explain the data distribution of strength of the silicate feature vs. [25]–[12] color (see fig. 8 and discussion in §4.4). Second, the models provide a unique variation of $\tau_F$ with [25]–[12] color. The correlations predicted for modified silicate $Q_{abs}$ disagree significantly with that displayed by the data listed in table 1, when $\tau_F$ is taken from analysis of dynamical properties of the wind (see §5.2).

The other possibility is to fill the region with tracks for mixed chemical compositions. For example,

– 10 –the figure shows that the track for a mixture of 20% silicate and 80% amorphous carbon grains passes right through the center of the distribution of IRAS objects. Indeed, the percentage of sources with distinct silicate features increases with distance from the track for pure amorphous carbon toward the one for pure silicate grains, as would be the case if the chemical composition varied continuously. Mixtures duplicate the success of the previous solution without its shortcomings, producing correlations of $\tau_F$ and silicate feature strength with [25]–[12] color in agreement with observations. For these reasons we feel that this possibility is more likely to be correct.

While mixtures of amorphous carbon and silicate grains properly explain the data, this possibility is somewhat controversial. Based on equilibrium models, it is widely believed that such mixtures are impossible and that dust grains should be either purely silicate or carbonaceous since the less abundant element is fully incorporated into CO (e.g. Gilman, 1969; Salpeter, 1974a; Gail, 1991); silicate grains are expected in oxygen stars, carbonaceous grains in carbon stars. However, a number of carbon stars do display silicate features (Little-Marenin, 1986; Willems & de Jong, 1986; Zuckerman & Dyck, 1986; Zuckerman & Aller, 1986; Evans, 1990; Skinner et al, 1990; Chan & Kwok, 1991; LeVan et al, 1992) while numerous oxygen stars display the SiC feature (Zuckerman & Aller, 1986; Skinner et al, 1990). Furthermore, OH maser emission has been detected in stars whose IR spectra give no indication of silicate grains (Chengalur et al. 1993). In fact, departures from equilibrium, such as molecule dissociation by shocks and accretion onto dust grains (Sedlmayr, 1988; Brown & Charnley 1990; Gail, 1991), could play a significant role in gas-dust chemistry, so the formation of dust grains with a mixed chemical composition cannot be excluded. Indeed, Skinner et al. (1990) conclude that the type of dust grains to condense when the photospheric [C]/[O] is close to unity is unpredictable. This could explain the lack of OH maser emission in a sample of stars believed to be oxygen rich (Eder et al, 1988; Gaylard, 1989; Lewis, 1989, 1992; Lewis et al, 1990; Chengalur et al, 1993). Although these stars are located in the same region of the color-color diagram as OH/IR stars and show silicate features, the detection rate of maser emission is less than $\sim 50\%$. If departures from equilibrium in the process of dust formation for stars with $[C]/[O] \gtrsim 1$ produced only $\sim 10\%$ silicate grains, these grains would dominate the infrared spectral shape, which would be identical to that of oxygen stars with pure silicate dust. Based on their infrared signature, traditionally these stars would be classified as oxygen stars even though the abundance of oxygenated molecules in their envelope material would be low, precluding OH maser action. Indeed, the detection rate for OH maser emission increases with the percentage of silicate grains in the winds inferred from analysis of their IR spectra (Hajian, 1994).

In summary, although controversial, mixtures provide a natural explanation for various observed characteristics of late-type stars, and the conflict with grain formation theory may not be as severe as is widely perceived. The possibility of mixed grain compositions in the winds around late-type stars deserves further scrutiny.

While most of the sources fall in the region predicted by the model calculations, the upper left portion of the diagram contains sources that cannot be fitted with any reasonable modification to the absorption coefficients. If we define

$$[60] - [25] < 0.46([25] - [12]) - 0.28 \qquad (4\text{-}5)$$

as the criterion for sources that can be adequately explained by our modeling, then this criterion is violated by 192 sources (13% of the sample). These sources display an excess of long wavelength emission and their nature can be further studied by considerations of colors based on the 100 $\mu$m flux, which were mostly ignored until now. The lower panel of figure 6 displays the [100]–[60] vs. [25]–[12] diagram, whose tracks show that the predicted [100]–[60] color for late-type stars should be nearly independent of both grain composition and $\tau_F$. The reason is that the envelopes are always optically thin at these wavelengths, even when $\tau_F$ is very large. While our models always produce [100]–[60] < −0.3, 52% of the IRAS sources in our sample have colors above this region, displaying a large excess of cool emission. Virtually all the sources that violate the criterion of equation 4-5 belong to this population.

In an attempt to explain this large discrepancy we were led to considerations of cirrus emission, the cool radiation that provides a background flux $C(\lambda)$ at the location of a point source. While $C(\lambda)$ is presumed negligible at 12, 25 and 60 $\mu$m, it can be significant at 100 $\mu$m, and the IRAS team has attempted to remove this contamination. Consequently, the listed 100 $\mu$m flux is presumed to be independent of this background emission. $C(100)$ can be estimated from the listed IRAS quantity $cirr3$, the surface brightness at 100 $\mu$m around the point source; it is simply $C(100) = 1.2 \times cirr3$ Jy, accounting for the beam size at 100 $\mu$m (note that $cirr3$ is listed in MJy ster$^{-1}$). Figure 7a displays the [100]–[60] color vs. $cirr3/F(60)$ for each source. If the listed $F(100)$ were indeed independent of the background cirrus emission then the figure would display a scatter diagram, and if all sources were late-type stars the scatter would be confined to a horizontal strip centered on $[100] - [60] \sim -0.5$. As the figure shows, this is indeed the case as long as $cirr3/F(60) \lesssim$ 1–5. However, sources with a larger value of $cirr3/F(60)$ display an essentially perfect correlation of $F(100)$ with cirrus flux. The reason for this tight correlation is as follows[2]: At long wavelengths, IRAS signal-to-noise ratio is inversely correlated with $cirr3/F(60)$. As $cirr3/F(60)$ increases, the signal-to-noise ratio decreases and the inherent uncertainty in the determination of $F(100)$ increases, causing the distribution to broaden toward both larger and smaller values of [100]–[60]. At the same time, the IRAS team choice of $F(100)$ threshold for listing in the point source catalogue introduced a correlation between the threshold and $cirr3$. The raising of the $F(100)$ threshold at large values of $cirr3$ removes the lower part of the resulting broadened distribution, leaving the rising strip displayed in the figure. This effect also produces a correlation between $F(60)$ and $F(100)/F(60)$, recently noted by Zuckerman (1993).

The tight correlation displayed in figure 7a indicates that the extraction of actual 100 $\mu$m point-source fluxes from the data is imperfect at large values of $cirr3/F(60)$. As a result, *long wavelength IRAS point source fluxes are unreliable when $cirr3 \gtrsim (1-5) \times F(60)$.* Indeed, the IRAS team was aware of these pitfalls and the IRAS Supplement (1988) issued cautionary notes that cirrus emission "can corrupt 100 $\mu$m, and occasionally 60 $\mu$m, measurements of point sources" (p. I-2 of the IRAS Supplement). If cirrus emission ever dominated also at 60 $\mu$m, the linearly rising portion of the diagram would stop at a value of $cirr3/F(60)$ corresponding to $C(100)/C(60)$. Since there are no sources with $cirr3/F(60) \gtrsim 250$, we can set an upper limit of 2.5 on the cirrus [100]–[60] color. If the cirrus spectrum follows a black-body at temperature $T$, this color implies $T \gtrsim 11$ K. For emission proportional to $\nu^\beta B_\nu(T)$, $T \gtrsim 10$ K for $\beta = 1$ and $T \gtrsim 9$ K for $\beta = 2$. These results are in agreement with the estimate by Knapp et al (1990) that $T \sim 20$ K.

A meaningful evaluation of our model results can be accomplished only after removal of the cirrus contaminated sources from the sample, a procedure involving some ambiguity. As is evident from figure 7a and the previous discussion, contaminated sources are those that have $cirr3/F(60) > a$, where $a$ is somewhere between 1 and 5. Therefore, we have constrained our sample according to this criterion and present the results in table 2, where cirrus removal was based on $a = 1$, 2 and 5. For each $a$ we list the total number of sources that remain in the sample and among those, the number and percentage of "bad" sources, those that cannot be fitted by our models. The third and fourth columns present the effect on the [60]–[25] color-color diagram, where "bad" sources are those that violate criterion 4-5. The fifth and sixth columns provide the same information for the [100]–[60] color-color diagram, where "bad" sources are those with [100]–[60] > −0.3. As is evident from the table, the effect of contamination removal on the

---

[2] We thank Tom Chester for his help in clarifying this issue.

[60]–[25] diagram is essentially independent of $a$. Removal with any of those three values shows that only about 4% of the remaining sources cannot be explained by our models. As could be expected, contamination removal from the [100]–[60] diagram is more sensitive to the value of $a$. Since $a = 1$ brings the percentage of "bad" sources to agreement in both color-color diagrams, it seems safe to assume that, with this value of $a$, cirrus contamination has been properly removed from our sample. However, cirrus removal with $a = 2$ doubles the size of the sample with only a slight increase in the percentage of "bad" sources in the [100]–[60] color-color diagram. Figure 7b presents the same color-color diagrams as figure 6, but with cirrus contamination removed using $a = 2$. The figure shows that among the handful of sources that cannot be fitted by our models there is a distinct group of 14 sources, marked with solid triangles. These sources were selected because they stand out as a concentration in the [100]–[60] diagram and turn out to mostly occupy a distinct region also in the [60]–[25] diagram. They are listed in table 3 and correspond to 5% of the sample of 292. All are optically bright and have rather low mass-loss rates, corresponding to extremely small $\tau_F$, still they display large excess of cool emission, which remains unexplained. Considering observational and model uncertainties and noise in the sample, essentially all other sources are well explained by our model predictions. That is, *all IRAS fluxes are properly explained by our models for 95% of all uncontaminated sources in the VH window.*

As table 2 shows, the initial IRAS sample in the VH window contained 192 sources whose [60]–[25] colors could not be explained by radiation pressure driven winds in steady state. To explain these sources, Willems & de Jong (1988) proposed a model of detached shells caused by time variations in the mass-loss rate. In this model, late-type stars follow a loop through the upper portion of the [60]–[25] — [25]–[12] diagram. Were this proposal to work, all sources without time varying mass-loss rates would have to be free of cirrus contamination while all sources whose mass-loss rates did vary with time were dominated by cirrus emission. This seems unlikely. Recently, Zuckerman (1993) also questioned the Willems & de Jong model on similar grounds. Still, observational evidence does exist for both time variations and detached shells in a number of sources. Olofsson et al. (1990) surveyed 89 bright C stars for CO emission and detected it in 65 of them, of which 3 were found to have a double-peaked profile. Although Olofsson et al. suggest that this profile is indicative of a geometrically thin, expanding shell, they point out that the observed CO emission need not imply time variations in the mass-loss rates. Furthermore, the Willems & de Jong model also predicts oxygen enhanced chemistry for the detached shells. This prompted Bujarrabal & Cernicharo (1994) to search for emission of oxygen rich molecules from the thin CO shells that Olofsson et al. discovered. This search produced negative results.

Another possible explanation for the thin shell CO emission and the large far-IR excess is that the shells mark the interface between winds and surrounding medium. A steadily blowing wind will pile-up the ambient gas into a shell after a time

$$t \simeq 10^4 \left(\frac{\dot{M}_{-6}}{n_2 v_{10}^3}\right)^{1/2} \text{ years} \tag{4-6}$$

(e.g. Steigman et al. 1975). Here $\dot{M}_{-6}$ is the mass-loss rate in units of $10^{-6}$ $M_\odot \text{yr}^{-1}$, $v_{10}$ is the expansion velocity in 10 km s$^{-1}$ and $n_2$ is the density of the ambient medium in $10^2$ cm$^{-3}$. The shell radius is

$$r \simeq 3 \times 10^{17} \left(\frac{\dot{M}_{-6}}{n_2 v_{10}}\right)^{1/2} \text{ cm,} \tag{4-7}$$

comparable to the sizes that Olofsson et al. find for the shells. It may be significant that two of the three shell sources are cirrus contaminated (the third one, U Ant, is one of the fourteen problematic sources listed in table 3), as cirrus emission can be expected to mark interstellar regions with elevated densities. Since the time scale for shell onset is comparable to evolutionary times for the mass-loss process, the shell and the currently observed wind can be expected to have different properties, such as velocity for example.

Our analysis shows that once cirrus contamination is properly removed, virtually all IRAS sources in the VH window can be explained as late-type stars. Therefore, we can reasonably assume that the same holds for the bulk of the contaminated sources too. Additional support for this proposal is provided by the detection of OH maser emission for sources in the VH-window. From the data of Chengalur et al (1993) we find that the detection rate ($\sim 38\%$) is essentially independent of the [100]–[60] color. A slight decrease in detection rate occurs when $F(25)$ decreases below $\sim 10$ Jy, consistent with a decrease below the OH sensitivity limit due to variability, as proposed by Chengalur et al. In addition, the average [100]–[60] color is the same ($\sim 0.6$) for both detections and non-detections, the result expected if all sources were selected from the same population.

Are other types of objects excluded? For another population to mimic the IRAS properties of late-type stars its members would have to have the same spectral shapes $f_\nu$ in the infrared. That could be achieved with the same radial variation of dust absorption coefficient, namely, the same function $g(x)$ (see equation 4-3). Although such a population has not yet been identified, this possibility cannot be ruled out.

### 4.4. LRS Spectral Classes

The LRS catalogue lists four spectral classes for "blue" sources, those whose flux decreases faster than $\lambda^{-1}$ in the 14–22 $\mu$m range. Featureless spectra are classified by an index $10 + n$, where $n$ ($1 \leq n \leq 9$) is proportional to the 8–13 $\mu$m spectral slope. Spectra with structure in the 9.7 $\mu$m region are classified by $20 + n$ when the feature is in emission and $30 + n$ for absorption. The index $n$ is then determined from the band strength at 9.7 $\mu$m, the ratio of observed flux to interpolated continuum. Spectra with the 11.3 $\mu$m SiC feature, always in emission, are classified as $40 + n$, and $n$ is again proportional to the appropriate band strength.

Since both colors and LRS indices are determined from flux ratios, a given spectral shape fully determines both. Therefore, for any model we can readily determine both the LRS class and colors. Van der Veen & Habing (1988) find that LRS classes occupy well defined regions in the color-color diagram (see figure 3b of their paper), and scaling provides a simple explanation for this finding: the spectral shape, including spectral features and colors, is uniquely determined by the dust composition and $\tau_F$. Figure 8 displays the variation of LRS class type with [25]–[12] color, comparing our model predictions with the data. Typical observational errors are $\sim 0.1$ in the [25]–[12] color. In each panel we present the color distribution for each LRS class, the average color for the class obtained from statistical analysis by van der Veen & Habing (1988), and the model results for relevant grain compositions. Except for SiC, for every grain type the [25]–[12] color increases monotonically with $\tau_F$, thus the class variation reflects the evolution of spectral shape with optical depth. It is important to note that the spread around each average evident in the figure gives an exaggerated, false impression of the actual spread in the data. Each data point in the figure corresponds to all the sources with that color and spectral class, and thus represents a large number of sources. This number is different for different points, decreasing significantly with distance from the class average. Therefore, the FWHM of the data distribution at every class is only about half the maximal spread seen in the figure. Note also the different [25]–[12] scales of the three panels.

The top panel corresponds to featureless, class $1n$ sources. Class 18 is a black-body spectrum ($\tau_F = 0$) that can be produced by all grains, but only carbonaceous grains can explain featureless spectra with smaller $n$ (unless silicate dust grains have very large radii, $> 2$ $\mu$m, which is unlikely). As is evident from the figure, the curve produced by amorphous carbon models is a reasonable fit for the averages of this class while the spread around the average is within the observational errors. Therefore, amorphous carbon adequately describes the data while pure graphite grains are ruled out. Since $n$ *decreases* when $\tau_F$ increases and class 18 corresponds to $\tau_F = 0$, neither grains can produce class 19 sources. The fluxes of these sources decrease with wavelength faster then $\lambda^{-4}$ for $\lambda > 8$ $\mu$m and could be attributed to either uncertainties in measurements or an additional component of the radiation field, stronger at 8 $\mu$m than at 13 $\mu$m (emission by PAHs in the dust grains is a possible candidate, Buss et al, 1991).

Classes $2n$ and $3n$ correspond to silicate dominated grains. As $\tau_F$ increases from 0, the [25]–[12] color increases from $-0.58$ and the 9.7 $\mu$m emission feature increases its strength, corresponding to class $2n$ sources with increasing $n$. When $\tau_F \sim 0.3$ (corresponding to $\tau_\nu(9.7) \sim 1$ and [25]–[12] $\sim -0.20$) the silicate feature reaches its maximum emission strength and $n$ begins to decrease with further increase of the [25]–[12] color. The silicate feature remains in emission so long as $\tau_F \lesssim 1$ ($\tau_\nu(9.7) \lesssim 3$), corresponding to [25]–[12] $\lesssim 0.2$. As is evident from the figure, each class $2n$ can correspond to two different values of $\tau_F$, i.e., two different [25]–[12] colors. The separation of these two branches exceeds significantly the observational errors ($\sim 0.1$). They properly bracket the observed distribution and their averages agree with those deduced from the data by van der Veen & Habing. However, the data points fill the entire area bounded by the pure silicate track rather than being confined to a band around the track whose width is the observational errors. Such a distribution can only be explained by mixed grain composition, as indicated by the plotted track for a 1:4 mixture of silicate and amorphous carbon. This provides additional support for the presence of mixed grain composition.

About 10% of class $2n$ sources display an additional minor peak at 11 $\mu$m, indicating the presence of crystalline olivine. The abundance of this component cannot exceed $\sim 20$–30% because higher abundances would produce an 11 $\mu$m peak more prominent than the 9.7 $\mu$m feature. Such a situation is observed in only 3 out of more than 100 class $2n$ sources that show the 11 $\mu$m feature.

With further increase in optical depth, $\tau_F \gtrsim 1$ and [25]–[12] $\gtrsim 0.2$, the silicate feature switches to absorption whose depth increases with $\tau_F$, i.e., with [25]–[12] color, corresponding to class $3n$ sources with increasing $n$. Although the trend of the variation of index $n$ with color predicted by our models is in agreement with the observations, the depth of the 9.7 $\mu$m feature is generally underestimated. However, the index $n$ of class $3n$ sources is determined from

$$n = -11.5 \log \frac{F(9.7)}{F_c(9.7)}, \tag{4-8}$$

where $F_c$ is the continuum interpolation across the absorption feature, a procedure that greatly magnifies small errors in the fitting. When the feature is very deep ($F_{9.7} \sim 0.2 F_c$, corresponding to $n = 38$), the flux at 9.7 $\mu$m is comparable to typical noise levels and is very sensitive to calibration errors (Neugebauer et al, 1986). Considering the uncertainties involved, our models produce reasonable explanations for classes $2n$ and $3n$.

Objects with SiC feature belong to classes $4n$ and $n$ increases with the strength of the 11.3 $\mu$m emission feature. The lower panel shows the model predictions for the variation of $n$ with color for two mixtures of SiC and amorphous carbon grains. The behavior of the curves is similar to that for spectra with Sil emission features (class $2n$) and the agreement with the data is satisfactory. There are various indications that classes above 46–47 can be discounted. Van der Veen & Habing (1988) found only 6 objects belonging to classes 48 and 49, and Chan & Kwok (1990) noticed that sources classified as 46–49 almost always have noisy spectra. Since $n > 7$ is excluded by the data we can set an upper limit of $\sim 20$–30% on the fraction of SiC grains in mixtures, in agreement with the limit deduced earlier from the color-color diagram. It may be also worthwhile to note that Griffin (1990) successfully fitted the spectrum of IRC+10216 with a mixture of 20% SiC and 80% amorphous carbon grains and that Lorenz-Martins & Lefevre (1993) found for a sample of nine stars that the percentage of SiC to amorphous carbon grains never exceeds 20%.

It should be noted that LRS classification can be problematic when not aided by complete radiative transfer modeling. For example, the LRS class of IRAS 14119-6453 has been designated 42, corresponding to the SiC 11.3 $\mu$m in emission. However, the lower left panel of figure 4 shows that the spectrum of this source is adequately fitted with pure silicate dust at an optical depth $\tau_F = 8$. The apparent peak around 11 $\mu$m is actually generated by the 9.7 $\mu$m silicate feature, reflecting the effect of its absorption at this optical depth. Similar mis-classifications can be noted for a number of other sources in the VH-window. It is also important to note that the LRS spectral classification is somewhat crude. For example, many class 14 and 15 sources, classified as featureless, actually have broad, weak emission features (e.g. Little-Marenin & Little, 1990). Similarly, the results we display are only meant to explain the overall behavior of the data. The addition of minor dust components that could improve the agreement between model calculations and data was not attempted. A lot more work will need to be done to see if the IRAS colors and the LRS classification of any individual source can be reproduced with the same grain composition.

### 4.5. Radial Profiles

The radial dependence of the problem is fully described in terms of the scaling variable $x = r/r_0$, where $r_0$ is the radius of the dust formation point. It varies according to $r_0 \propto L_*^{1/2} T_0^{-2}$, where $T_0$ is the temperature of dust condensation (see NE). Figure 9 displays the variation of $\nu F_\nu / F$ with radial distance for silicate grains and $L_4 = 2$ at $\lambda = 2$, 12, 25 and 60 $\mu$m, and $\tau_F = 0.1$, 1.0 and 10.0. The radiation at 2 $\mu$m is generated almost exclusively by the star and is attenuated by dust absorption during passage through the envelope. As $\tau_F$ increases, the fraction of the total flux emerging at this wavelength diminishes and the star is obscured by the envelope. At the three IRAS wavelengths the flux is dominated by dust emission and it increases with $\tau_F$ in envelopes with moderate optical depths. In thick envelopes, self-absorption by the outer layers is evident at 12 $\mu$m. The radius at which the flux reaches its final value increases with $\lambda$, thus the observed envelope size increases with wavelength, as observed by Harvey et al (1991). At 12 $\mu$m the flux reaches its final value at $r \lesssim 10^{16}$ cm. *Spectral features indicative of silicate and SiC grains are formed in the inner parts of the envelopes.*

We find that the radial dependence of the dust temperature $T_d$ is described adequately by the familiar power law dependence expected for optically thin envelopes with $Q_\nu^a \propto \lambda^{-\beta}$,

$$T_d(r) \propto r^{-\frac{2}{4+\beta}} \tag{4-9}$$

(see e.g. Harvey et al. and references therein). The index $\beta$ is 2.0 for graphite and 1.5 for silicate, amorphous carbon and SiC grains, respectively. Although the spectral index of $T_d$ decreases slightly in optically thick envelopes, this effect can be neglected. The dust temperature decreases to $\sim 20$–30 K at $r \sim 10^{18}$ cm, defining the cutoff radius for our numerical integration. This value is in agreement with the conclusion of Rowan-Robinson et al. (1986) that the ratio of an envelope inner to outer radius is $\sim 0.001$, and with the size measurement of the IRC+10216 envelope in the CO $J = 2$–1 line by Knapp et al. (1982).

Beginning with Leung (1975), radiative transfer calculations for stars with dusty envelopes have usually employed the density distribution $\rho(r) \propto r^{-2}$. Our models show this approximation to be less than adequate since sizeable departures do exist in the inner regions of the envelopes, $r \lesssim 10 r_0$, where most of the acceleration takes place and the density falls off faster than $r^{-2}$. We find these departures to have a noticeable effect on the radiative transfer problem. For example, Chan and Kwok (1990) found significantly different tracks in the color-color diagram for models that employ a constant velocity outflow and those that simulate accelerated winds. The latter agree better with our results. Recently, Groenewegen (1994) concluded from comparison of radiative transfer models with IR data that the density falls off faster than $r^{-2}$, an effect he attributes to variation in the mass-loss rate. Our models, which properly incorporate the dynamics, show that, because of the acceleration, this density variation is actually produced at a constant mass-loss rate.

## 5. OBSERVATIONAL IMPLICATIONS

### 5.1. Determination of $\tau_F$ from Observed Spectra

Thanks to scaling, all the spectral properties are fully determined by $\tau_F$, so a reliable determination of this quantity from the data is important. In principle, any spectral quantity can be used for this purpose. Van der Veen (1989) proposed to determine $\tau_F$ from the $F(25)/F(12)$ ratio, and employed this method for OH/IR stars where he found a power law correlation between these two quantities. However, modeling by Justtanont & Tielens (1992) shows that a single power law cannot describe all the data when optically thin envelopes are included. Our modeling supports this finding. Furthermore, we find that the correlation of $F(25)/F(12)$ and $\tau_F$ depends rather strongly on the grain composition; different grains produce correlations that differ significantly from each other, so $\tau_F$ cannot be determined if the grain composition is not known.

From our model calculations we find that the most adequate quantity for this purpose is the ratio $(\nu F_\nu)_{60}/F$ because its dependence on $\tau_F$ is almost independent of the grain chemical composition. This correlation is displayed in figure 10, which shows that all the dust grains considered in our models produce almost the same curve; different curves differ by less than a factor of 2. A simple power law fit to the detailed model results is

$$\tau_F = 130 \frac{(\nu F_\nu)_{60}}{F}, \tag{5-1}$$

valid for all grains and $(\nu F_\nu)_{60}/F \gtrsim 4 \times 10^{-4}$ (corresponding to $\tau_F > 0.05$). Optical depths obtained from this relation can be considered reliable to within factor 2–3, typically. In applying this relation, care must be taken to avoid cirrus contaminated sources. The reason $(\nu F_\nu)_{60}/F$ provides such a simple, useful correlation is that the envelopes are optically thin at 60 $\mu$m even when $\tau_F$ is very large. The usefulness of this relation is enhanced at high optical depths ($\tau_F \gtrsim 2$, $(\nu F_\nu)_{60}/F \gtrsim 0.01$), where the correlation displayed in figure 3 can be invoked to eliminate the bolometric flux in terms of the [25]–[12] color. On the other hand, at lower optical depths the correlation is so steep as to become useless for $\tau_F < 0.05$ [at $\tau_F = 0.05$, $(\nu F_\nu)_{60}/F$ hardly differs from its value at $\tau_F = 0$]. At these low optical depths we find that $\tau_F$ can be determined instead from the emission features at 9.7 $\mu$m and 11.3 $\mu$m when they are present. If the feature strength at wavelength $\lambda_0$ is defined as

$$A_{\lambda_0} = 2.5 \log \frac{F(\lambda_0)}{F_c}, \tag{5-2}$$

where $F_c$ is the interpolated continuum across the feature, we find that whenever one of these features is present and $(\nu F_\nu)_{60}/F < 0.001$,

$$\tau_F \propto A_{\lambda_0}^{1.4}. \tag{5-3}$$

Here the proportionality constant is, respectively, 0.1, 0.2, 0.3 and 0.2 for pure silicate, 1:1 and 1:4 mixtures of silicate and amorphous carbon, and 1:4 mixture of SiC and amorphous carbon.

## 5.2. Determination of $\dot{M}$

Equation 3-3 provides a straightforward method to determine the mass-loss rate when $L_*$, $\tau_F$, $v$ and $\Gamma$ are known. The outflow velocity $v$ can be determined directly from OH and CO measurements and various methods to estimate the overall luminosity $L_*$ are also available. The methods we just introduced can be used to determine $\tau_F$, and the only unknown quantity is the gravitational correction $\Gamma$. This correction can be more conveniently related to directly observable quantities by introducing the velocity

$$v_\Gamma = \tau_F \frac{L_*}{\dot{M}\Gamma c} = GM_* \int_{r_0}^{\infty} \frac{dr}{r^2 v(r)}. \tag{5-4}$$

Then the mass-loss rate can be determined from

$$\dot{M}_{-5} = 20 \frac{\tau_F L_4}{v + v_\Gamma}, \tag{5-5}$$

where $\dot{M}_{-5} = \dot{M}/10^{-5} M_\odot$ yr$^{-1}$, and $v$ and $v_\Gamma$ are in km s$^{-1}$. The integral on the right-hand side in the definition of $v_\Gamma$ is proportional to $r_0^{-1}$ and the dust condensation radius $r_0$ is itself proportional to $L_*^{0.5}$. While $M_*$ is known only for a limited number of objects, based on data from van der Veen (1989) we find that $M_*/M_\odot \simeq 0.8 L_4$. This leads to

$$v_\Gamma \propto L_4^{0.5}, \tag{5-6}$$

where the proportionality constant is about 18 km s$^{-1}$ for carbonaceous grains, 30 km s$^{-1}$ for Si-based grains.

Mass-loss rates obtained from this new method agree on average within a factor of 2 with determinations by other methods, when available, for the stars listed in table 1. In particular, Skinner & Whitmore (1988a,b) recently obtained phenomenological correlations between $\dot{M}$ and $A_{\lambda_0} \times F(12)$ for samples of stars with silicate and SiC features. The relations we find lead to almost identical correlations if the scatter in $v$ and $L_*$ is neglected, using $v = 15$ km s$^{-1}$ and $L_4 = 1$ for all stars.

## 6. DISCUSSION

Our modeling shows that steady-state outflows driven by radiation pressure on dust grains adequately describe the surroundings of late-type stars. Both the infrared emission and outflow dynamics are well accounted for. Even the highest mass loss rates measured thus far can be supported by radiation pressure, and all spectral features and colors are adequately explained when cirrus emission is taken into account. For 95% of the sources there is no need to augment the model with either time dependent effects or an additional component such as a detached shell. In all likelihood, essentially all IRAS sources in the VH-window are late-type stars.

Since our models assume steady state they can only describe the time-averaged behavior of the outflows. This is an adequate description as long as time variability occurs only on time scales shorter than the averaging time. The radial profiles of figure 9 show that most of the observed flux is formed within $r \sim 10^{16}$ cm. With typical outflow velocities of $\sim 10$ km s$^{-1}$, this distance is covered in $\sim 300$ years, the relevant time scale for averaging. Although all late-type stars display pulsational or irregular variability, such variations are characterized by a few years at most, much shorter than the averaging time. Thus the assumption of steady state seems justified. Additional time dependence can be introduced by chemical evolution of the stellar atmosphere, possibly leading to gradients in the grain chemical composition. Though our models do not include such an effect, they adequately describe the envelopes as long as chemical gradients occur only at $r \gtrsim 10^{16}$ cm. The reason is that the grain composition in that region is largely irrelevant since all grains behave similarly at the cool temperatures and long wavelengths relevant there. And since spatial gradients with length scale $\lesssim 10^{16}$ cm imply temporal variations with time scales $\lesssim 100$ years, the durations of phases not covered by our models do not exceed $\sim 100$ years. If typical lifetimes in the mass-losing phase are $\sim 10^4$–$10^5$ years, only $\sim 0.1$–1% of all sources cannot be described by our models, in agreement with the statistics presented in table 2. Evolutionary effects characterized by time scales longer than $\sim 100$ years can be incorporated into our models as adiabatic changes of the relevant sources.

Another important ingredient of our models is the assumption of spherical symmetry. Small departures from sphericity, such as a slight elongation, should not affect our results significantly. However, our models do not apply in the case of major deviations from spherical symmetry such as disk geometry or strong clumpiness when their characteristic size scales are less than $\sim 10^{16}$ cm. The success of our models indicates that such deviations may not be important for the majority of late-type stars. It is worthwhile to note that OH maser profiles suggest that less than 20% of sources can deviate from spherical symmetry (Chengalur et al, 1993).

All the absorption coefficients employed in our modeling are tabulated in the literature and have been deduced from detailed theoretical and experimental studies. Still, uncertainties exist. In the case of silicate, we employ the long wavelength fall-off typical of amorphous grains. Both this property and the modification of the silicate feature strengths based on Simpson's (1991) analysis significantly improve the agreement between model results and observations.

For simplicity, a single size was used here for all grains. A distribution of grain sizes would affect both the dynamics through the dust drift and the radiative transfer through the absorption coefficient. The effect on the dust drift was discussed in NE, where it was found negligible. Similarly, NE have shown that because $Q/a$ is independent of $a$ when $2\pi a/\lambda \ll 1$, the absorption coefficient is independent of size distribution as long as the fraction of mass incorporated in dust is fixed. Therefore, grain size variations cannot have a significant impact on radiative transfer at the long wavelengths considered here.

Our models show that infrared observations alone suffice to deduce important parameters of the outflow. Scaling provides a method to determine the optical depth, which can then be utilized to determine the mass-loss rate. The grain chemical composition can also be deduced. Roughly, the [25]–[12] color is correlated with optical depth, and for a given value of this color the [60]–[25] color is indicative of the grain chemical composition. Though the spread of this color is rather small, together with LRS class variation with color the composition can be determined with some confidence. Graphite grains seem to be ruled out, as does an SiC abundance in excess of $\sim 20$–30%. The silicate feature provides a clear indication of the presence of silicate grains, but the actual percentage of this component in a mixture cannot be determined with great accuracy as there is little difference among abundances larger then $\sim 10$%. Conclusions drawn from spectral shapes can be quite misleading when not supported by full radiative transfer modeling. The evolution of a spectral feature with optical depth can be easily mistaken for a different behavior. Examples are provided by the spectra presented in the lower left panel of figure 4 (see §4.4) and in figure 5 (see §4.2).

Late-type stars provide a highly successful, self-consistent explanation for all IRAS sources in the VH-window. Still, as long as it is not known what are the intrinsic IR fluxes and colors generated by other types of objects it cannot be claimed with certainty that this explanation is unique. To tackle this problem, we plan to study in a future publication the general properties of astronomical objects that control the characteristics of their IR emission and determine their location in the IRAS color-color diagrams.

We thank Nathan Netzer for valuable comments about the computer code, John Mathis for providing us with optical properties of amorphous carbon, Bob Stencel for pointing out the potential significance of cirrus emission, Mike Hauser and Tom Chester for extensive discussions concerning the IRAS data analysis and Murray Lewis for the data of OH maser detections and non-detections in IRAS sources. This research has made use of the SIMBAD database, operated at CDS, Strasbourg, France, and the ADS database. Support by NSF grant AST-9016810 and the Center for Computational Sciences of the University of Kentucky is gratefully acknowledged.

## A. OUTFLOW VELOCITY IN RADIATIVELY DRIVEN WINDS

Retaining only the radiation pressure force, the radial component of the equation of motion of a circumstellar shell is

$$\rho \frac{dv}{dt} = \frac{1}{c}\rho \chi_F F, \tag{A1}$$

where $d/dt$ is convective derivative and $\chi_F$ is flux-averaged opacity per unit mass. Employing mass conservation, in steady state ($\partial/\partial t = 0$) this equation becomes

$$\dot{M} dv = \frac{L}{c} d\tau_F, \tag{A2}$$

where $L = 4\pi r^2 F$ is the shell luminosity at radius $r$ and $d\tau_F = \rho \chi_F dr$. This is simply a statement of momentum flux conservation: the change in mechanical momentum flux across a shell of thickness $dr$ is equal to the amount absorbed from the momentum flux of the radiation field, $L/c$, available at $r$. The shell luminosity can be obtained from energy flux conservation

$$L + \tfrac{1}{2}\dot{M}v^2 = L_* + \tfrac{1}{2}\dot{M}v_0^2, \tag{A3}$$

where $v_0$ is the initial outflow velocity. It is convenient to introduce a velocity $v_L$ through

$$L_* = \tfrac{1}{2}\dot{M}v_L^2. \tag{A4}$$

That is, $v_L$ is an upper bound on the outflow velocity, obtained in the limit in which the wind captures all the stellar radiation and fully converts it to mechanical energy. Since $v_L = 3490(L_4/\dot{M}_{-5})^{1/2}$ km s$^{-1}$, in all cases of interest $v_L \gg v_0$ and the solution of eq. A2 is

$$v = v_0 + v_L \frac{\exp(\tau_F v_L/c) - 1}{\exp(\tau_F v_L/c) + 1}. \tag{A5}$$

The limit velocity $v_L$ is approached when $\tau_F v_L/c \gg 1$. Note that the condition for full conversion of radiative luminosity into mechanical luminosity is not $\tau_F > 1$ but rather $\tau_F v_L/c > 1$. The opposite limit, $\tau_F v_L/c \ll 1$, yields the familiar result

$$v - v_0 = \tau_F \frac{v_L^2}{2c} = \tau_F \frac{L_*}{\dot{M}c}, \tag{A6}$$

same as eq. A2, but with finite differences replacing differentials. Indeed, when $\tau_F v_L/c \ll 1$, $v \ll v_L$ so $\tfrac{1}{2}\dot{M}v^2 \ll L_*$ and $L \simeq L_*$, the shell luminosity is constant (the fraction of radiative flux converted into mechanical energy flux is negligible). Since $\tau_F v_L/c = 0.01\,\tau_F(L_4/\dot{M}_{-5})^{1/2}$, this is the relevant limit in all cases of interest.

Although the validity region of eq. A6 is $\tau_F v_L/c \ll 1$, it has often been misstated as $\tau_F < 1$ because of alleged momentum conservation. This misunderstanding seems to arise from the confusion between momentum and momentum flux, and differences between the velocities of matter and radiation. Equation A6 is the result of convective differentiation, moving with the material. The quantity $\dot{M}\Delta v$ is the change in mechanical momentum during the time the matter crosses a shell of thickness $\Delta r$ and must be compared to the amount of radiative momentum emitted by the star *during the same time*. Because of the difference in velocities, the distance covered by radiation during the material shell crossing time is $c/v$ times larger. The radiative momentum available to impart force to the shell of thickness $\Delta r$ fills the much larger volume of thickness $(c/v)\Delta r$ because all the radiation contained in this volume has streamed through the shell while the matter has crossed it. Therefore, the effective radiative momentum flux is $(L_*/c) \times (c/v) = L_*/v$ and eq. A6 shows that only the fraction $\tau_F v/c \ll 1$ of this available flux is actually captured by the shell and converted into mechanical momentum flux. It is for this same reason that full conversion of radiative luminosity into mechanical luminosity occurs only when $\tau_F > c/v_L$, not when $\tau_F > 1$.

– 21 –

Fig. 1.— Absorption efficiency for 0.05 $\mu$m spheres made of astronomical silicate, crystalline olivine, graphite, SiC and amorphous carbon. In the case of silicate, the thin dotted line is from Draine & Lee (1984), the thick solid one incorporates modifications indicated by the data analysis of Simpson (1991) and a long wavelength behavior corresponding to amorphous structure ($Q \propto \lambda^{-1.5}$). The modified $Q_{abs}$ is the one used in the model calculations.

Fig. 2.— Domain of radiatively driven winds: mass outflows can be supported by radiation pressure on dust grains only in the $\Gamma > 1$ region of the corresponding chemical composition. Sources marked by stars with CRL numbers have $\dot{M}v > L_*/c$ (data are from Jura, 1986, assuming $M_* = 10\ M_\odot$). The numbers in parenthesis are the corresponding values of $\tau_F$ with $\Gamma$ calculated for amorphous carbon grains.

Fig. 3.— Test of correlations predicted by scaling (§4.1). Data points are for the stars in table 1. Open squares mark O-stars, solid circles C-stars. Curves describe the model predictions for various grain compositions, as marked.

Fig. 4.— Model spectra (normalized $\nu F_\nu$) for astronomical silicate grains for a sequence of $\tau_F$. Solid lines present the spectral shape of the emergent radiation, the sum of two components — the dust emission, indicated in dotted lines, and the emerging stellar radiation, the dashed lines. The model prediction for the [25]–[12] color is indicated. In each panel, the inset displays the measured spectrum of an actual source whose IRAS number is marked above the top left corner. The measured [25]–[12] color of the source is listed in the inset in large type.

Fig. 5.— Panels on the left are from Little-Marenin & Little (1990), displaying spectra representative of the Sil, Sil+ and Sil++ classes they proposed. Each panel plots the observed IRAS spectrum (normalized $\nu F_\nu$) of the indicated source, a 2500 K black-body distribution matched to the stellar spectrum at 8 $\mu$m, and the difference spectrum (observed minus black-body continuum); the latter was presumed to be proportional to the dust absorption coefficient. Panels on the right display results of model calculations. Dot-dashed lines are the stellar 2500 K black-body radiation and dashed lines are this radiation as it emerges after attenuation by the dust. Dotted lines are the dust emission and full lines are the sum of this component and the emergent stellar radiation, the observed spectra. In all of these models the absorption coefficient is the same, corresponding to pure astronomical silicate grains; the only difference is the model optical depth $\tau_F$, as indicated. The lower right panel, $\tau_F = 0.03$, displays also model results when 20% crystalline olivine is included, plotted with thin solid lines.

Fig. 6.— IRAS color-color diagrams of all sources in the VH-window (see §4.3), displayed as dots, with quality 3 fluxes at 12, 25 and 60 $\mu$m, and quality 2 or 3 at 100 $\mu$m. The top panel is the [60]–[25] vs. [25]–[12] diagram. Solid lines represent theoretical sequences for different grain compositions, as marked. In all mixtures, the abundances ratio of the first to second component is 1:4. The common origin of all tracks ($[25] - [12] = -0.58$, $[60] - [25] = -0.76$) corresponds to $\tau_F = 0$. Distance from that common origin along each track increases with $\tau_F$ and diamond marks indicate the locations of $\tau_F = 0.1$, 1.0 and 10.0. Bottom panel is the [100]–[60] vs. [25]–[12] diagram, where the order of the tracks from bottom to top around $[25]–[12] \sim 0$ is graphite, Oli. + Sil, Sil, Sil + am.C, am.C and SiC + am.C. On each of these tracks, the locations of diamond marks would be at the same [25]–[12] colors as for the corresponding track in the upper panel.

Fig. 7.— (a) [100]–[60] color vs. $cirr3/F(60)$ for all sources, where $cirr3$ is the IRAS surface brightness at 100 $\mu$m around the point source. Sources for which only a lower limit on $cirr3$ is listed, about 20 of them, are omitted. (b) The color-color diagrams of figure 6, removing sources with $cirr3/F(60) > 2$.

Fig. 8.— Distributions of LRS classes vs. [25]–[12] color (see §4.4). IRAS sources are displayed as dots and for each LRS class an open circle marks the average color from the statistical analysis by van der Veen & Habing (1988). Typical observational uncertainties are $\sim 0.1$ for the colors. Note the different scales of the [25]–[12] color in the three panels. In the middle panel, sources that display the Sil 9.7 $\mu$m feature in absorption are denoted by crosses and their average colors by open squares. Lines indicate the model predictions for grain compositions as marked. (a) Class $10 + n$ sources, corresponding to featureless spectra. (b) Classes $20 + n$ and $30 + n$, corresponding to spectra that display the 9.7 $\mu$m feature in emission and absorption, respectively. (c) Class $40 + n$ sources, displaying the 11.3 $\mu$m feature. In the cases of class $20 + n$ and $40 + n$, the van der Veen & Habing results correspond to averages of the two branches of the theoretical curves (see text).

Fig. 9.— Radial variation of $\nu F_\nu/F$ at 2, 12, 25 and 60 $\mu$m for envelopes with astronomical silicate grains when the stellar luminosity is $2 \times 10^4$ $L_\odot$. Each panel corresponds to a different value of $\tau_F$, as indicated.

Fig. 10.— Variation of $\tau_F$, the flux averaged optical depth, with $(\nu F_\nu)_{60}/F$ in models with various grain compositions, as marked. Also displayed is the indicated power law fit to the results.